\documentclass[]{pasj00}

\begin{document}
\SetRunningHead{T. Yamaguchi et al.}{Entry Dispersion Analysis for Hayabusa}

\title{Entry Dispersion Analysis for the Hayabusa Spacecraft using Ground Based Optical Observation \thanks{Based on data collected at Subaru Telescope, which is operated by the National Astronomical Observatory of Japan; and MegaPrime/MegaCam, a joint project of CFHT and CEA/DAPNIA, at the Canada-France-Hawaii Telescope (CFHT) which is operated by the National Research Council (NRC) of Canada, the Institut National des Science de l'Univers of the Centre National de la Recherche Scientifique (CNRS) of France, and the University of Hawaii.}}

\author{Tomohiro \textsc{Yamaguchi} %
  \thanks{JSPS Research Fellow}}
\affil{The Graduate University for Advanced Studies}
\email{yamaguchi.tomohiro@jaxa.jp}

\author{Makoto \textsc{Yoshikawa}}
\affil{Japan Aerospace Exploration Agency}\email{yoshikawa.makoto@jaxa.jp}

\author{Masafumi \textsc{Yagi}}
\affil{National Astronomical Observatory of Japan}\email{YAGI.Masafumi@nao.ac.jp}

\and

\author{David J. \textsc{Tholen}}
\affil{University of Hawaii}\email{tholen@IfA.Hawaii.Edu}


%

\KeyWords{astrometry --- celestial mechanics --- Hayabusa --- orbit determination --- space vehicles} 

\maketitle

\begin{abstract}
Hayabusa asteroid explorer successfully released the sample capsule to Australia on June 13, 2010. Since the Earth reentry phase of sample return was critical, many backup plans for predicting the landing location were prepared. This paper investigates the reentry dispersion using ground based optical observation as a backup observation for radiometric observation. Several scenarios are calculated and compared for the reentry phase of the Hayabusa to evaluate the navigation accuracy of the ground-based observation. The optical observation doesn't require any active reaction from a spacecraft, thus these results show that optical observations could be a steady backup strategy even if a spacecraft had some trouble. We also evaluate the landing dispersion of the Hayabusa only with the optical observation. 
\end{abstract}

\section{Introduction}
Recently, many scientists have planned sample return missions to achieve further understandings for planets and small bodies. The Discovery-class mission Stardust of the National Aeronautics and Space Administration (NASA) was launched in 1999 to collect the dust from comet Wild-2 \citep{key0}. After seven-year journey, Stardust finally released its capsule with cometary and interstellar dust particles into Utah Test and Training Range. 

Japanese Space Exploration Agency (JAXA) launched the Hayabusa spacecraft in 2003 to accomplish a sample return from the near-Earth asteroid (25143) Itokawa. The Hayabusa arrived at the target asteroid in 2005 with its ion thrusters and the Earth gravity assist. The Hayabusa was originally scheduled to return to the Earth in 2007, however the actual Earth entry was carried out in 2010 due to the unexpected trouble at the sampling phase in 2005. Finally, the Hayabusa capsule has successfully landed to Woomera Prohibited Area (WPA) on June 13th, 2010. Since, the Earth entry phase of the Hayabusa spacecraft was a critical event, several backup observations were prepared.

The Hayabusa spacecraft was mainly navigated using radiometric observations. Range, Doppler and Delta differential one-way range (DDOR) observable were provided by the communication between the spacecraft and ground stations. However, these radiometric observations always require an active reaction from a spacecraft. If a spacecraft had some trouble in its communication module, these precise measurements were no longer available. On the other hand, ground-based optical observations are completely passive and don't require any active reaction from a spacecraft. Therefore optical observation is a steady backup strategy for a spacecraft trouble. 

In this paper, we evaluate the orbit determination of the Hayabusa reentry phase using ground-based observations as a backup strategy of the nominal radiometric observation. Several scenarios were calculated and compared for the reentry phase of the Hayabusa to evaluate the navigation accuracy of the ground-based observations. The real radiometric tracking data and optical observation data are evaluated. This evaluation could be a good index for future reentry missions. Comparison of these results reveals the importance of ground based observations as a backup strategy for Earth entry events. 

In section 2, we briefly describe the Hayabusa observation, both optical and radiometric measurements. The detail of observatories and their conditions are presented. The analysis method and conditions for the entry dispersion analysis are described in section~3. Several cases are determined in order to compare the dispersion ellipses with and without the optical observation. Next, the results and comparison of the analysis are investigated using the radiometric and the optical observations in section~4. Finally, we summarize our conclusion in section~5.

\section{Observation}
\label{sec:obs}
\begin{figure}
  \begin{center}
    \FigureFile(80mm,80mm){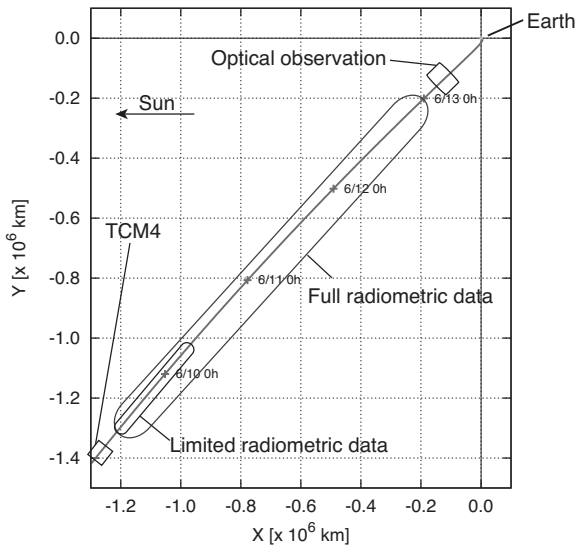}
  \end{center}
  \caption{Hayabusa Earth reentry trajectory  with Earth center, Sun-Earth fixed rotating frame. Hayabusa position at the radiometric and the optical observation are also described. A description about the limited and full radiometric data is written in section 3.}
  \label{fig:hayabusa_obs}
\end{figure}

After the five series of trajectory correction maneuvers, the Hayabusa spacecraft was guided into WPA of Australia \citep{key7}. The Earth approaching trajectory is described in figure~\ref{fig:hayabusa_obs} on rotating frame. The Sun is always located in $-X$ direction in this figure, and the Hayabusa spacecraft approached the Earth from dayside.

\subsection{Ground based optical observation}
\begin{figure}
  \begin{center}
    \FigureFile(80mm,80mm){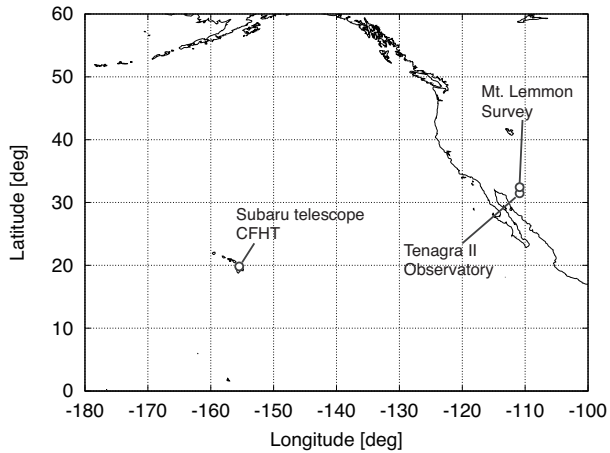}
  \end{center}
  \caption{This location of the observatories for Hayabusa observation.}\label{fig:site}
\end{figure}

Hayabusa was observed by four ground-based optical observatories located in Arizona and Hawaii (figure \ref{fig:site}). Table~\ref{tab:obs_comp} shows the complete data of the Hayabusa optical observations used in this study. Many observatories in Japan also tried to observe Hayabusa from June 11 to 13, though none of them succeeded due to poor weather conditions.

\begin{table}
	\caption{Hayabusa optical observations. The time is expressed as hours from June 13.0, 2010. The site code 926, G96, SBR, CFH correspond to Tenagra II observatory (\timeform{110D.879} W, \timeform{31D.462} N, 1309 m), Mt. Lemmon Survey (\timeform{110D.789} W, \timeform{32D.443} N, 2789 m), Subaru telescope (\timeform{155D.4761} W, \timeform{19D.8255} N, 4162 m) and Canada-France-Hawaii telescope (\timeform{155D.4688} W, \timeform{19D.8252} N, 4204 m), respectively.}\label{tab:obs_comp}
	\begin{center}
	\begin{tabular}{c|ll|c}
    		\hline
		Time [UTC] & RA [HMS] & DEC [DMS] & Site  \\
		\hline
3.6878 &08 54 02.12 & +28 00 18.1   &   926 \\
3.7445 &08 54 03.66 & +27 59 08.2   &   926 \\
3.8009 &08 54 05.28 & +27 57 57.0   &   926 \\
4.2230 &08 54 24.44 & +27 47 01.2   &   G96 \\
4.2372 &08 54 25.02 & +27 46 43.2   &   G96 \\
4.2516 &08 54 25.77 & +27 46 22.0   &   G96 \\
4.3308 &08 54 29.23 & +27 44 33.4   &   G96 \\
4.3778 &08 54 31.42 & +27 43 26.5   &   G96 \\
4.3884 &08 54 31.96 & +27 43 13.1   &   G96 \\
4.3997 &08 54 32.68 & +27 42 55.8   &   G96 \\
4.4090 &08 54 33.04 & +27 42 42.3   &   G96 \\
4.4340 &08 54 34.37 & +27 42 05.1   &   G96 \\
4.4549 &08 54 35.38 & +27 41 35.8   &   G96 \\
4.4654 &08 54 35.91 & +27 41 22.4   &   G96 \\
5.9945 &08 56 36.63 & +28 02 30.9   &   SBR \\
6.0182 &08 56 37.09 & +28 01 53.1   &   SBR \\
6.0295 &08 56 37.31 & +28 01 35.2   &   SBR \\
6.0516 &08 56 37.76 & +28 01 00.1   &   SBR \\
6.0617 &08 56 37.97 & +28 00 44.0   &   SBR \\
6.0732 &08 56 38.23 & +28 00 25.0   &   SBR \\
6.0850 &08 56 38.48 & +28 00 06.5   &   SBR \\
6.1066 &08 56 38.98 & +27 59 31.2   &   SBR \\
6.1722 &08 56 40.529& +27 57 43.36  &   CFH \\
6.1967 &08 56 41.178& +27 57 02.29  &   CFH \\
6.2215 &08 56 41.848& +27 56 20.94  &   CFH \\
\hline
    \end{tabular}
  \end{center}
\end{table}

\begin{table}
  \caption{Hayabusa observation by the Subaru Telescope.}\label{tab:subaru}
  \begin{center}
    \begin{tabular}{ll}
      \hline
	Camera & Suprime-Cam\\
        Pixel size & 0.202 arcsec\\
        Seeing & 0.6-0.7 arcsec\\
	Exposure time & 5.0 sec $\times$ 11 shots \\
	Filter & V-band(W-J-V; 5500 $\AA$) \\
	Tracking mode & Sidereal tracking \\
	RA, Dec & 08:56:34, +27:57:33 \\
	Time  & Twilight time on June 13, 2010 \\
      \hline
    \end{tabular}
  \end{center}
\end{table}

\begin{figure}
  \begin{center}
    \FigureFile(80mm,80mm){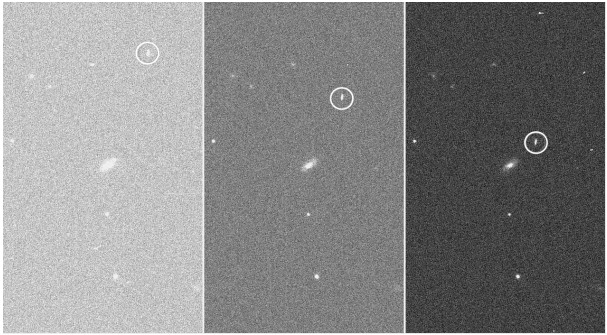}
  \end{center}
  \caption{Three of Hayabusa images taken with the Subaru telescope. 
North is up and east is left. As they were observed in sidereal tracking 
mode, the Hayabusa image (indicated with a circle) elongates in 
the direction of the movement.
}
\label{fig:subaru_img}
\end{figure}

\subsubsection{Subaru observation}

The Hayabusa spacecraft was observed with Suprime-Cam \citep{key1}
mounted on the Subaru telescope in the evening of June 11 - 13 (UT).
Since the Hayabusa spacecraft approached the Earth from the dayside,
the window for the optical observation was small.
The Hayabusa was detected only in June 13 data. The successful
observation by the Subaru telescope is summarized in
table~\ref{tab:subaru}. The pointing was set so that the trajectory
will be in a single readout channel of a chip (chip 5) of Suprime-Cam
for an easy quicklook. The seeing size was 0.6-0.7 arcsec. 
The brightness of the Hayabusa was estimated to be V=19.8 mag in 
AB magnitude. The photometric calibration was performed using SDSS DR7
\citep{key10} stars in the observed field with 
a conversion of SDSS magnitude to Suprime-Cam V-band magnitude.

The time system of the Suprime-Cam is synchronized to UTC using a network
time protocol(NTP).  The typical shift of the Suprime-Cam time from
UTC is less than 2$\mu$s. However, the Suprime-Cam FITS header of the
end of the exposure (UT-END keyword) has uncertainty of -1 to 0
seconds because of the communication delay from the camera to the
control computer. The time of start of the exposure (UT-STR) is much
more accurate as the start command sent from the control computer to
the camera. The mean time of the observation was therefore calculated
from UT-STR and the half of the exposure time (EXPTIME). Then we add
0.6 seconds to correct for the effect of the shutter movement, for the
shutter moves in 1.2 seconds and the chip 5 is near the center.  As
the Suprime-Cam has not implemented the information about the shutter
movement, either the shutter curtain moved from the left to the right,
or from the right to the left in the field, the shutter movement would
make the time uncertainty by about 0.1 second in our
observation. Three data are removed because large delay between the
computer and the camera communication by some trouble is recorded.

The data were reduced in a standard manner using nekosoft
\citep{key2}; bias subtraction, flat fielding, distortion correction 
and sky subtraction. We did not apply PSF equalization process
among the exposures. The centroid of Hayabusa was also measured with the
software. WCS has been calibrated in each exposure using WCStools
\citep{key3} and USNO-B1.0 catalog \citep{key4}. Nine or ten stars
were used in each exposure and the RMS residual was 0.17$\pm$0.02 arcsec in the best-fit solutions.



\subsubsection{CFHT observation}
Observations of Hayabusa were attempted on both June 12 and 13 with the 3.6-m Canada-France-Hawaii Telescope (CFHT) on Mauna Kea. The instrument Megaprime was utilized along with an r filter. The camera consists of 36 Marconi/EEV 2048$\times$4612 pixel CCDs covering a 1$\times$1 deg field of view; a single CCD covers a 6.4$\times$14.4 arcmin field.  The ephemeris of Hayabusa was known well enough to place the target near the center of chip 22 of the camera, which is located immediately south of the optical axis. The exposure times were 40 sec in all cases, and non-sidereal tracking at the expected rates of motion for the spacecraft were used on both nights.  

Three exposures were taken on June 13, and the spacecraft was easily detected with a signal-to-noise ratio in excess of 70.  The measured brightness of the spacecraft was R=19.0, using nearby astrometric reference stars from the USNO-B1.0 catalog to determine the photometric zero point.

\begin{figure}
  \begin{center}
    \FigureFile(80mm,80mm){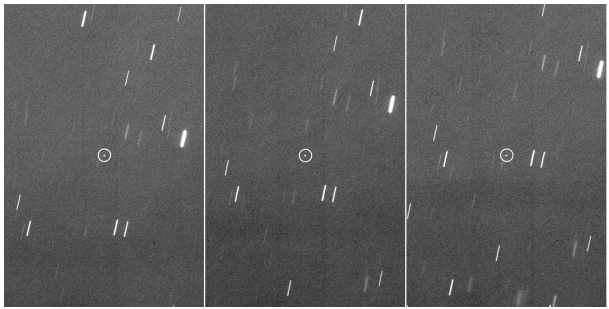}
  \end{center}
  \caption{R-band images of Hayabusa taken with the Canada-France-Hawaii Telescope on 2010 June 13 at (left to right) 6:10:00, 6:11:28, and 6:12:57 UT.  North is up and east is left.  The stars are trailed by 18 arcsec during the 40 sec exposures tracked at the spacecraft's rates of motion.  The measured magnitude of Hayabusa is about R=19.0 with SNR in excess of 70.}\label{fig:cfht_img}
\end{figure}

The astrometric solutions utilized a quadratic field distortion model and a minimum of 62 reference stars from the USNO-B1.0 catalog.  The three independent solutions for the three exposures showed RMS residuals of 0.15 to 0.17 arcsec for the reference sources, thus the contribution from the astrometric solution to the overall positional uncertainty was only about 0.02 arcsec.  Centroiding error on the spacecraft contributed another 0.01 arcsec.  A +5.0 sec correction to the exposure start times in the FITS headers was applied to compensate for the known delay between when the clock is read and when the camera shutter actually opens.  The 0.3 sec jitter in this delay contributed about 0.12 arcsec of astrometric uncertainty in declination and about 0.02 arcsec in right ascension. Because the reference stars were all trailed by 49 (binned) pixels due to the non-sidereal tracking mode used for the exposures, a model consisting of a trapezoid in the trailed direction and a Gaussian in the orthogonal direction was used to find the centroid of each reference star's image.  A Gaussian model was used for the image of Hayabusa itself, which was not trailed (see figure~\ref{fig:cfht_img}).

\begin{table}
  \caption{Hayabusa observation by theCFHT.}\label{tab:cfht}
  \begin{center}
    \begin{tabular}{ll}
      \hline
	Camera & Megacam \\
	Pixel ssize &  0.187 arcsec,  \\
	& binned 2$\times$2 to 0.374 arcsec \\
	Seeing & 0.7 arcsec \\
	Exposure time & 40.1 sec $\times$ 3 shots \\
	Filter & R-band (r.MP9601) \\
	Tracking mode & non-sidereal at expected  \\
	& spacecraft rates \\
	Queue observer & A. Draginda \\
	Queue coord. & G. Morrison \\
      \hline
    \end{tabular}
  \end{center}
\end{table}

\subsubsection{Other observations}
Hayabusa was also observed from two sites in Arizona. One site is Mt. Lemmon Survey operated by A. Gibbs. No filters were used and UCAC2 \citep{key9}  star catalog was used for the reference with sidereal tracking. Exposure time was  30 seconds. 

Another site is Tenagra II observatory. P. Holvorcem observed Hayabusa with Apogee AP8 camera (back-illuminated SITe chip, 1 K $\times$ 1 K). The reference catalog was USNO A2.0\footnote{http://www.usno.navy.mil/USNO/astrometry/optical-IR-prod/icas/the-pmm/readme.v20} and tracking mode was non-sidereal. Unfortunately, the detail information is not available because the original data were lost.

\subsubsection{Brightness of Hayabusa}
The Subaru data taken on June 12 were 1 minutes exposure and 
2 shots of 2 minutes exposures in V-band.
The estimated detection limit in each exposure was about 22.3 mag,
As the tracking mode of Subaru observation was sidereal, 
the limiting magnitude is shallower than that of CFHT observation. 

Two exposures were taken on June 12 with CFHT, but the target was not detected in either of them, or in a stacked version of both exposures, which has an effective exposure time of 80 sec.
Using the inverse-square law, the topocentric distance on June 12 suggests a brightness of V=22.7, which should have been detectable.  The difference in phase angle between the two nights (133.7 deg on June 12, 132.4 deg on June 13) shouldn't have been enough to cause an appreciable brightness difference (an asteroidal phase function would predict a 0.14 mag brightening between the two nights).  We therefore attribute the non-detection on June 12 to the spacecraft's orientation with respect to the observer.  An object with large, flat surfaces and specular reflection could easily have a large variation of brightness.

We also stacked both CFHT 40 second exposures and the three Subaru 60/120/120 second exposures to detect the position of Hayabusa at June 12. However, we cannot detect the signal and it seems that Hayabusa was fainter than expected based on the June 13 data corrected for distance.

\subsection{Radiometric observable}
The continuous radiometric tracking data were provided by JAXA Usuda deep space center and NASA deep space network. We could receive the radiometric observable 24 hours using these network. Actual reentry was navigated by radiometric data. These measurements provide 2-way Doppler and range observable. These observables are sensitive with respect to the line of sight direction, but not the tangential direction. Also these observables require reaction of a spacecraft. A spacecraft needs to receive the radio signal from ground and transmits back to a ground station. Therefore, these observables are only available if the spacecraft's communications system is in a healthy condition.

\section{Analysis and its conditions}
In this section, we compare the dispersion ellipse with and without the optical observations in order to evaluate the impact of optical observation. The orbit determination and entry dispersion analysis are investigated to evaluate the impact of the ground based observation of the Hayabusa. The trajectory of the Hayabusa are estimated using several data sets and evaluated with a dispersion ellipse on the B-plane \citep{key5} and the Earth's surface. The estimation method is conventional weighted least squares fit. We adopted the weights of the 2-way Doppler and range observation as 0.5 mm s$^{-1}$, 10 m, respectively. The weights of the optical observation varies with the observatories. The observations of Tenagra observatory, Mt. Lemmon survey, Subaru telescope and CFHT are weighted as 0.6, 1.0, 0.3 and 0.3 arcsec, respectively. These values are validated using the residuals of the orbit determination process. A planetary perturbation using Jet Propulsion Laboratory ephemeris DE423\footnote{http://ssd.jpl.nasa.gov/?ephemerides} is considered for the trajectory propagation. The solar radiation pressure is considered using the cannonball model \citep{key6}. The Earth orientation model for terrestrial to celestial coordinate transformation is compliant with IAU 2000A CIO based\footnote{http://www.iers.org}. 

\begin{figure}
  \begin{center}
    \FigureFile(80mm,80mm){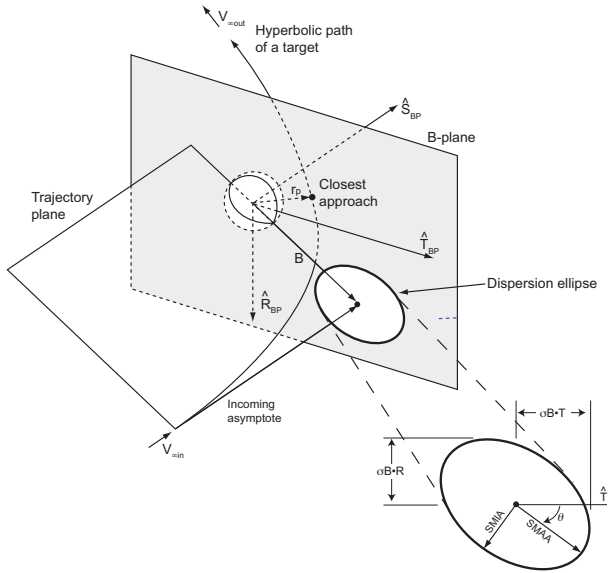}
  \end{center}
  \caption{Definition of B-plane coordinates}
  \label{fig:bplane}
\end{figure}

B-plane is the useful plane to design the targeting condition of entry and flyby. The plane is normal to the incoming hyperbolic velocity (figure~\ref{fig:bplane}). The origin of the B-plane coordinate system is the Earth center and the horizontal axis (B.T) is parallel to the equator plane of the Earth. The dispersion ellipse on the ground are calculated without considering the atmospheric effect, because the purpose of this study is to compare the dispersion ellipse with and without optical observations and describe the impact of ground based observations.

Three data sets are prepared for the observations and summarized in table~\ref{tab:obs_data}. Data set A is the all optical observation data taken by four observatories. Data set B is the limited version of the radiometric data which assumes that the Hayabusa spacecraft had an unexpected issue on June 10. Data set C is the nominal case of the radiometric data which includes all the radiometric data until June 13. Four cases are analyzed using the combination of the data sets and summarized in table~\ref{tab:cases}. Case 1 investigates the dispersion for the limited case and compare with case 2 to understand the effect of the optical observation. The difference of case 1 and 2 is the availability of the optical observations, therefore the difference of the dispersion ellipse describes the impact of the optical observations for reentry object navigation. Case 3 describes the nominal dispersion of the Hayabusa mission. The effect of the tracking arc for radiometric measurement is presented comparing case 1 and 3. Case 4 shows the dispersion only with the optical observations and this case corresponds to the test case for the Earth impact prediction of near Earth objects. 

\begin{table}
  \caption{Observation data.}\label{tab:obs_data}
  \begin{center}
    \begin{tabular}{lll}
    	\hline
	Data & Type & Time [UTC] \\
      \hline
      	A & Optical  & 6/13 3:41 \\
	&  observation & - 6:13  \\
	\hline
	B & Limited  & 6/9 11:00  \\
	   & radiometric data & - 6/10 6:30 \\ 
	\hline
	C & Full  & 6/9 11:00  \\
	   & radiometric data & - 6/13 0:00 \\ 
      \hline
    \end{tabular}
  \end{center}
\end{table}

\begin{table}
  \caption{Analysis cases.}\label{tab:cases}
  \begin{center}
    \begin{tabular}{ccl}
    	\hline
	Case & Observation & Comments \\
	&  data & \\
      \hline
      	1 & B & Some issue happen  \\
	 & & in the spacecraft \\
	\hline
	2 & A, B & Follow up observation  \\
	& & by ground-based telescope \\ 
	\hline
	3 & C & Nominal case  \\
	& &  (No trouble in the spacecraft)\\
	\hline
	4 & A & Optical observation only \\
      \hline
    \end{tabular}
  \end{center}
\end{table}


\section{Results and Discussion}
\begin{figure}
  \begin{center}
    \FigureFile(80mm,80mm){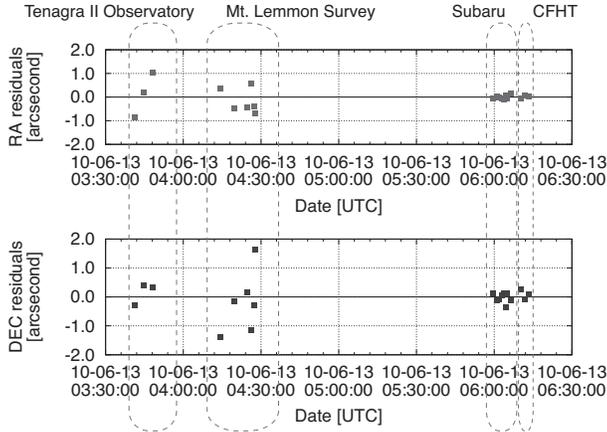}
  \end{center}
  \caption{RADEC residuals of case 4}
  \label{fig:resi_case4}
\end{figure}

The results of the orbit determination (OD) and the dispersion ellipses are investigated in this section. The post-fit residuals of case 4 are described in figure~\ref{fig:resi_case4}. Five observations of Mt. Lemmon Survey are rejected due to its large residuals. The observations of Subaru telescope and Canada France Hawaii Telescope (CFHT) are quite stable and all the observations fits within 0.5 arcsecond. The rms of 2-way Doppler and range data are 0.064~mm s$^{-1}$ and 0.30~m, respectively. The standard deviations of the position and velocity vector at the OD epoch are summarized in table~\ref{tab:solutions}. The impact of the optical observations is found in the difference between case 1 and 2. Especially, the uncertainties for the velocity dramatically decrease. It would be due to the extension of the tracking arc duration by the optical observation. However, the OD solution using the full radiometric observation (case 3) is much better than the hybrid case (case 2). Since case 4 have only 2.5 hours of optical observation, the solution has a large uncertainties along both position and velocity vector. The main uncertainty is along the velocity direction, because a optical imaging has no information along the line of sight direction.

\begin{table}
	\caption{Uncertainties for each analysis (epoch 2010/6/9 6:04 UTC)}\label{tab:solutions}
	\begin{center}
	\begin{tabular}{l|cccc}
    		\hline
		Case & 1 & 2 & 3 & 4 \\
	 	\hline
      		Radiometric  & 0.812 & 0.812  & 3.542 &  N/A \\
		arc [days] & &  & & \\
		\hline
      		Optical  & N/A &  0.106 & N/A &  0.106 \\
		arc [days] & &  & & \\
		\hline
		Position standard  & 1.602 & 0.439 & 0.161 & 6263.655 \\
		deviation [km] & & & & \\
		\hline
		Velocity standard  & 4.331 & 0.172 & 0.064 & 1866.316 \\
		deviation [cm s$^{-1}$] & & & & \\
		 \hline
    \end{tabular}
  \end{center}
\end{table}

\begin{figure}
  \begin{center}
    \FigureFile(80mm,80mm){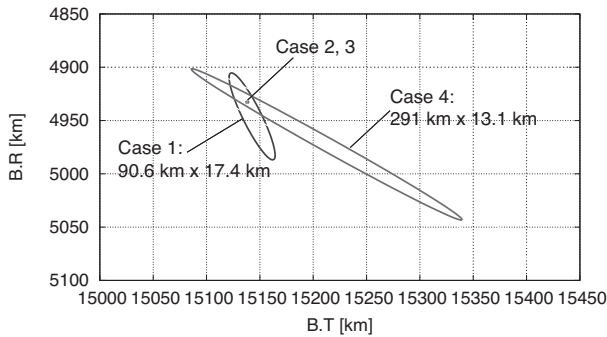}
  \end{center}
  \caption{3$\sigma$ B-plane dispersion ellipse }
  \label{fig:b_disp_all}
\end{figure}

\begin{figure}
  \begin{center}
    \FigureFile(80mm,80mm){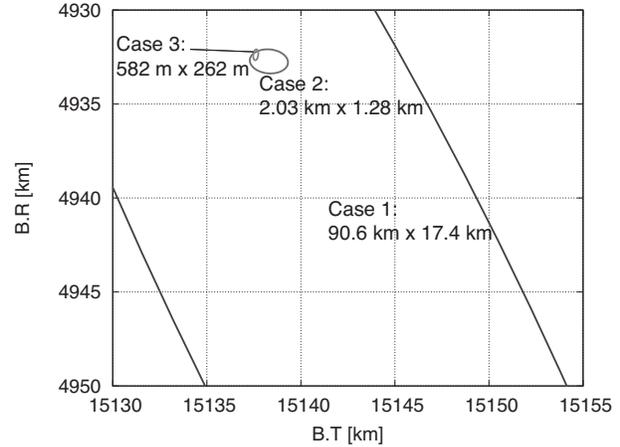}
  \end{center}
  \caption{3$\sigma$ B-plane dispersion ellipse (expand view)}
  \label{fig:b_disp_ex}
\end{figure}

The 3 sigma dispersion ellipse on B-plane is described in figure~\ref{fig:b_disp_all} and \ref{fig:b_disp_ex}. The major axis of case 4 is 291 km and it looks small compare with the uncertainty of the OD epoch, because B-plane is orthogonal to the hyperbolic infinite velocity and the main uncertainty is along the velocity direction. The dramatic improvement on the uncertainties by the optical observation are found comparing the ellipse of case 1 and 2. The size of the ellipse becomes about 1/600 of the original ellipse and the mean value becomes much closer to the value of case 3. It is natural that the ellipse of case 3 is the smallest in these cases, just because this case used the better quality and longer duration of observation data.

\begin{figure}
  \begin{center}
    \FigureFile(80mm,80mm){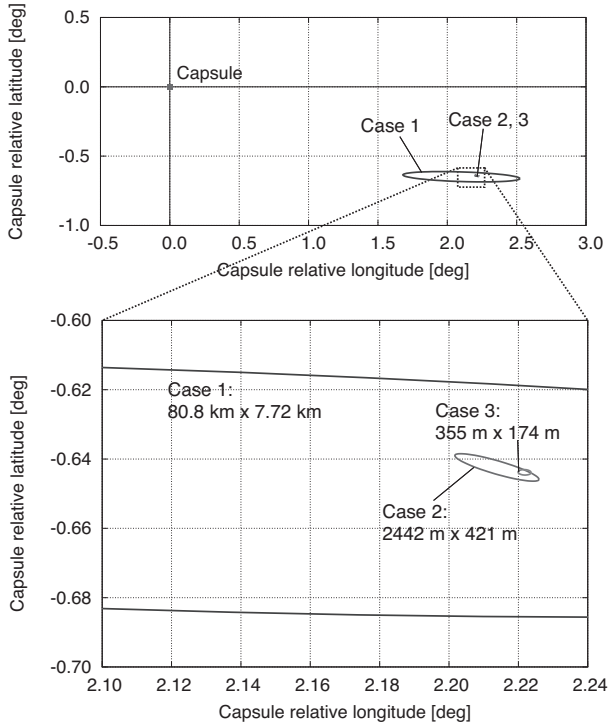}
  \end{center}
  \caption{3$\sigma$ landing dispersion ellipse relative to the landing site of the Hayabusa sample capsule}
  \label{fig:g_disp_all}
\end{figure}

\begin{figure}
  \begin{center}
    \FigureFile(75mm,80mm){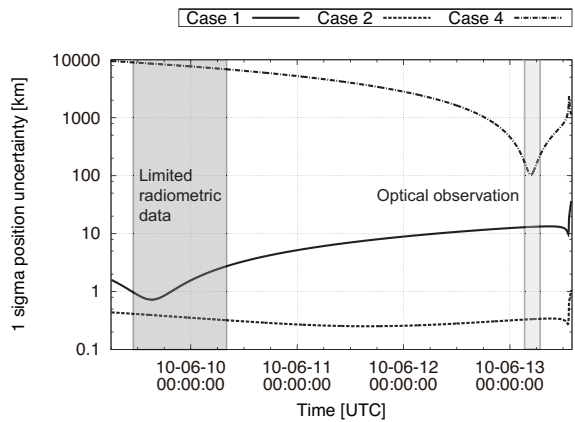}
  \end{center}
  \caption{Position dispersion history of the Earth approaching phase}
  \label{fig:cov}
\end{figure}

Figure~\ref{fig:g_disp_all} depicts the landing dispersion ellipse for case 1, 2 and 3 with the actual landing site of Hayabusa capsule. Since we haven't considered the atmospheric effect, there is a difference in the landing ellipse and the capsule landing site. Comparing the case 1 and 2, it is found that the major axis reduces more than one order due to the followup optical observation. The ground-based optical observation significantly improves the entry dispersion ellipse. This is because the observations were just before the entry and sensitivity of the optical observation to the trajectory is perpendicular to that of radiometric observable. The position dispersion history for case 1, 2 and 4 are shown in figure~\ref{fig:cov} with the observation time. The figure shows the relationship between the uncertainty and the observation. These results show that the optical observation becomes a strong backup strategy for sample return missions. 

\begin{figure}
  \begin{center}
    \FigureFile(80mm,80mm){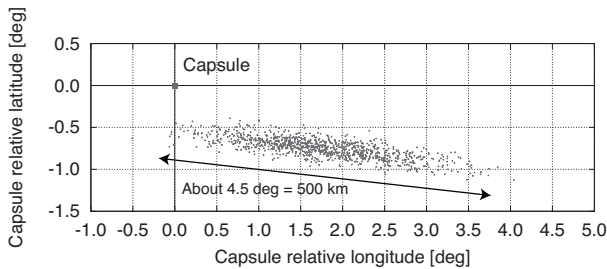}
  \end{center}
  \caption{Landing dispersion ellipse of case 4. The coordinates of the plots are relative to the actual landing site of the Hayabusa capsule.}
  \label{fig:g_disp_case4}
\end{figure}

Linear covariance method can break down if the position uncertainty grows too large. A linear assumption becomes less adaptive if uncertainties increase. In this study, the uncertainties of case 4 are too large to analyze the landing dispersion with linear covariance method. The landing dispersion ellipse of case 4 is described in figure \ref{fig:g_disp_case4} with the final landing location of Hayabusa sample capsule. This dispersion ellipse is the results of Monte Carlo simulation with 1000 particles. The terminate condition of the analysis was 0 km altitude of WGS-84 ellipsoid model. Consequence of large uncertainty along the velocity direction, the landing time uncertainty becomes 87.4 sec (1$\sigma$). The landing site predict with about 500 km accuracy, despite the fact that we only have the optical observations. The main difference between the actual capsule landing site and the predicted site is due to the atmospheric effect, however systematic biases on the star catalogs could affect the prediction. 

\begin{figure}
  \begin{center}
    \FigureFile(80mm,80mm){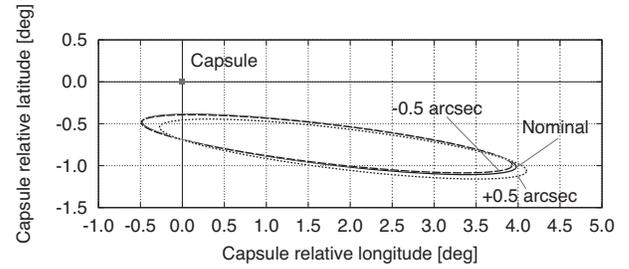}
  \end{center}
  \caption{Landing dispersion ellipse of case 4 with assuming star catalog biases in the declination. The biases are considered in Subaru and CFHT observations. The ellipses are calculated from the Monte Carlo simulation with 1000 particles.}
  \label{fig:bias}
\end{figure}

The USNO A2.0 and USNO B1.0 is known to have bias to 2MASS catalog, particularly in declination, and a bias as large as a half arcsecond is quite possible \citep{key8}. The landing error ellipse are depicted considering the $\pm$ 0.5 arcsec declination biases on the Subaru and CFHT observation in figure~\ref{fig:bias}. It is found that the star catalog biases is negligible in this situation. It is because the Hayabusa was close to the observatories at the operation and the impact of the errors in astrometric angular observations are limited. 

\section{Conclusion}
This paper investigated the entry dispersion analysis for the Hayabusa spacecraft using ground-based optical observations. The dispersion ellipse with and without optical observation showed the significant impact of the ground-based observations to the landing footprint. This results validated that a optical observation is a strong backup strategy for Earth reentry missions. 

The entry dispersion analysis only with optical observation described the rare test case for the Earth entry object. The landing location was predicted with 4.5 deg along the longitude using Monte Carlo analysis.

\bigskip

The authors would like to acknowledge Dr. T. Ohnishi of Fujitsu Ltd. for providing the radiometric tracking data of Hayabusa. We also thank to the staff of Subaru telescope and CFHT for supporting the observation of Hayabusa. We are grateful to A. Gibbs of Mt. Lemmon Survey and P. Holvorcem of Tenagra II observatory, who provide the Hayabusa observation. Part of the data analysis were carried out in the common use data analysis computer system at the Astronomy Data Center (ADC) of the National Astronomical Observatory of Japan. We would like to acknowledge an anonymous reviewer to clarify some points in this paper.  Finally, we are sincerely grateful to all the observers who tried to observe Hayabusa reentry. This work was supported by Grant-in-Aid for Japan Society for the Promotion of Science (JSPS) Fellows. 



\begin{thebibliography}{}

\bibitem[Abazajian(2009)]{key10}
Abazajian et al. 2009, \apjs, 182, 543

\bibitem[Chesley et al.(2010)]{key8}
{Chesley}, S.~R., {Baer}, J., {Monet}, D.~G.,  2010, Icarus, 210, 158


\bibitem[Desai et al.(2008)]{key0}
Desai, P. N., et al., 2008, J. Spacecraft and Rockets, 45, 1262

\bibitem[Kawaguchi et al.(2010)]{key7}
Kawaguchi, J., Yamada, T., Kuninaka, H., 2010, in Proc. of 61st International Astronautical Congress, IAC-10-A3.5.1


\bibitem[Mink(2002)]{key3}
Mink, D.J. 2002, in ASP Conf. Proc. 281. ADASS XI, 
ed. D.A. Bohlender, D. Durand, \& T.H. Handley (San Francisco: ASP), 169 

\bibitem[Miyazaki et al.(2002)]{key1}
Miyazaki, S., et al. \pasj, 2002, 54, 833

\bibitem[Monet et al.(2003)]{key4}
Monet, D.G. et al. 2003, \aj, 125, 984

\bibitem[Montenbruck \& Gill(2000)]{key6}
Montenbruck, O., Gill, E.\ 2000, Satellite Orbits (Springer, Netherland) 79


\bibitem[Portock(2000)]{key5}
Portock, B. M. 2000, in Proc. of Astrodynamic Specialist Conference, 507


\bibitem[Yagi et al.(2002)]{key2}
Yagi, M., Kashikawa, N., Sekiguchi, M., Doi, M., Yasuda, N., 
Shimasaku, K., and Okamura, S., 2002, \aj, 123, 66


\bibitem[Zacharias et al.(2004)]{key9}
Zacharias, N., Urban, S. E., Zacharias, M. I., Wycoff, G. L., 
Hall, D. M., Monet, D. G., Rafferty, T. J., 2004, \aj,127, 3043


\end{thebibliography}
\end{document}